\documentstyle[preprint,prl,aps,epsfig,floats]{revtex}
\tightenlines
\begin{document}
\draft
\preprint{
\vbox{\halign{&##\hfil         \cr
        & IPPP/01/04           \cr
        & DCPT/01/08           \cr
        & hep-ph/0102130       \cr
        & February 2001        \cr
        }}}

\title{Fragmentation Functions for Lepton Pairs}

\author{Eric Braaten}
\address{Physics Department, Ohio State University, Columbus, OH 43210, USA}

\author{Jungil Lee}
\address{Deutsches Elektronen-Synchrotron DESY, 
        D-22603 Hamburg, Germany}

\maketitle
\begin{abstract}
We calculate the fragmentation function for a light quark to decay into 
a lepton pair to leading order in the QCD coupling constant.
In the formal definition of the fragmentation function,
a QED phase must be included in the eikonal factor 
to guarantee QED gauge invariance.  
We find that the longitudinal polarization fraction is a 
decreasing function of the factorization scale, 
in accord with the intuitive expectation that the virtual photon 
should behave more and more like a real photon
as the transverse momomentum of the fragmenting quark increases.

\smallskip
\noindent
PACS numbers: 13.85.Qk, 12.38.-t, 13.87.Fh, 13.88.+e
\end{abstract}

\vfill \eject

The QCD factorization theorems for inclusive single-particle production
\cite{theorems} guarantee that the differential cross section at
sufficiently large transverse momentum $P_T$ has the approximate scaling
behavior $d\sigma / dP_T^2 \sim 1/P_T^4$.  The scaling part of the differential
cross section for producing the particle $H$ with momentum $P$ has the form
\begin{equation}
d\sigma[H(P)] \;=\;
\sum_i \int_0^1 dz \, d \hat \sigma[i(P/z),\mu] \,
D_{i \to H}(z,\mu) \;,
\label{fact}
\end{equation}
where the sum is over partons $i$
and $d \hat \sigma$ is the differential cross
section for producing an on-shell parton $i$ with momentum $P/z$.  
The fragmentation function $D_{i \to H}$ gives the probability for a
virtual parton with invariant mass of order $\mu$ to decay 
into a state that includes a particle $H$ with the fraction $z$ 
of the longitudinal momentum of the parton. 
The $\mu$-dependence of the fragmentation functions is governed by
Altarelli-Parisi evolution equations whose kernels can be calculated using
perturbation theory.  Large logarithms of $z \mu/P_T$ in $d \hat\sigma$
can be summed up by taking $\mu$ of order $P_T/z$.

The factorization formula (\ref{fact}) can serve as an operational definition
of the fragmentation functions.  However, the fragmentation functions can also
be given formal definitions as vacuum-to-vacuum matrix elements of operators
that involve projections onto asymptotic states that include the particle $H$. 
Curci, Furmanski, and Petronzio defined them as matrix elements of bilocal
operators in the light-cone gauge \cite{Curci-Furmanski-Petronzio}.  
Collins and Soper introduced gauge-invariant definitions of the 
fragmentation functions \cite{Collins-Soper}.  
They are defined as matrix elements of nonlocal
gauge-invariant operators that involve a path-ordered exponential of the gluon
field called an eikonal factor: 
\begin{equation}
{\cal E} = {\cal P} \exp \left( i g \int dl \, n^\mu A_\mu^a T^a \right) \,,
\label{eikonal-1}
\end{equation}
where $T^a$ is a generator of the appropriate representation of the gauge
group.  In the light-cone gauge $n^\mu A_\mu^a = 0$, the eikonal factor
collapses to 1 and the definition reduces to that of Curci, Furmanski, and
Petronzio.  

In the case of a light hadron $H$, the fragmentation function $D_{i \to H}$ 
cannot be calculated using perturbative methods.  
However, once $D_{i \to H} (z,\mu_0)$ has been measured as a function of $z$ 
at some initial scale $\mu_0$,
the Altarelli-Parisi equations can be used to evolve it to other scales $\mu$. 
An analysis of the fragmentation functions for light hadrons has recently been
used to obtain a high-precision determination of the QCD coupling constant
\cite{Kniehl-Kramer-Poetter}.

In the case of a heavy quarkonium state $H$, 
the fragmentation functions can be calculated using perturbation theory 
up to a few nonperturbative constants.  The NRQCD factorization formalism
\cite{B-B-L} can be used to express the fragmentation function in the form
\begin{equation}
D_{i \to H}(z,\mu) \;=\; 
\sum_n d_{i \to (Q \bar Q)_n}(z,\mu) \langle O_n^H \rangle\;,
\end{equation}
where the sum is over color and angular momentum states of a $Q \bar Q$ pair 
and the constants $\langle O_n^H \rangle$ are called NRQCD matrix elements.
The coefficients $d_{i \to (Q \bar Q)_n}$ of the NRQCD matrix elements can be
calculated using perturbation theory.  The first explicit calculations were the
coefficients of $\langle O_1(^1S_0) \rangle$ in the fragmentation functions for
$g \to \eta_c$ and $c \to \eta_c$ and the coefficients of 
$\langle O_1(^3 S_1) \rangle$ in the fragmentation functions for 
$g \to J/\psi$ and $c \to J/\psi$ \cite{Braaten-Yuan,B-C-Y}. 
They were calculated by Braaten, Cheung, and Yuan using the operational
definition of the fragmentation functions provided by the factorization theorem
(\ref{fact}).  The first explicit calculations using the formal
definition of the fragmentation functions were carried out by Ma \cite{Ma}.  
The formal definition is particularly useful for calculating the 
fragmentation functions beyond leading order in $\alpha_s$ 
\cite{Braaten-Lee:NLO}.

The factorization theorems for inclusive single-hadron production can be
applied equally well to photon production and to lepton-pair production.  
They imply that at sufficiently large $Q_T$, where $Q_T$ is the 
transverse momentum of the photon or lepton pair, 
the differential cross section has the
approximate scaling behavior $d \sigma / dQ_T^2 \sim \ 1/Q_T^4$.  
The scaling part of the differential cross section has the form 
(\ref{fact}).  At leading order in the QED interaction, 
only a photon term needs to be added to the sum over partons $i$.  
The fragmentation functions $D_{q \to \gamma}$ and
$D_{g \to \gamma}$ for real photons are nonperturbative, 
and only their evolution with $\mu$ is calculable in perturbation theory.  
The same is true of the fragmentation functions for a lepton pair 
with invariant mass $Q$ comparable to the scale $\Lambda$ 
associated with nonperturbative effects in QCD.  
On the other hand, if $Q$ is much larger than $\Lambda$, 
the fragmentation functions $D_{q \to \ell^+\ell^-}$ 
and $D_{g \to \ell^+\ell^-}$
are completely calculable using perturbative QCD.
Qiu and Zhang have recently introduced formal definitions for 
``virtual photon fragmentation functions" 
$D_{q \to \gamma^*}$ and $D_{g \to \gamma^*}$ that are 
equivalent to fragmentation functions for lepton pairs \cite{Qiu-Zhang}. 

In this paper, we calculate the fragmentation function 
$D_{q \to \ell^+ \ell^-}$ for a quark to decay into a lepton pair 
with large invariant mass at leading order in QCD perturbation theory.  
We calculate the fragmentation function using both dimensional 
regularization and an upper limit on the invariant mass of the 
fragmenting quark.  We find that the fragmentation function defined by
dimensional regularization has unphysical behavior 
except at asymptotic values of the factorization scale.
The fragmentation function defined by the upper limit 
on the invariant mass gives a longitudinal polarization 
fraction that decreases as the factorization scale increases,
in accord with the intuitive expectation 
that the virtual photon should behave more and more like a real photon 
as the transverse momentum of the fragmenting parton increases.

\begin{figure}
\begin{center}
\epsfig{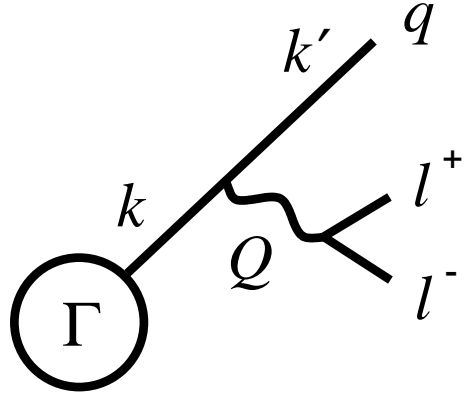}
\end{center}
\caption{
The class of Feynman diagrams that in the light-cone
gauge for QED gives the contribution to the
production of lepton pairs associated with fragmentation 
of the light quark $q$ at leading order in $\alpha_s$. 
}
\label{fig:operational}
\end{figure}

The fragmentation function for $q \to \ell^+ \ell^-$ can be calculated using
the operational definition provided by the factorization theorem, 
as in Refs. \cite{Braaten-Yuan,B-C-Y}.  
The Feynman diagrams for producing a lepton pair 
with large transverse momentum $Q_T$ include some with the topology
shown in Fig.~\ref{fig:operational}, 
where the blob represents the parts of the diagram that
involve the production of the virtual quark.  The terms in the
matrix element that corresponding to the fragmentation 
of a light quark with electric charge $e_q$ have the form
\begin{equation}
{\cal M} \;=\;
{1 \over k^2} \bar u(k') ( i e_q e \gamma_\mu ) \Gamma \,  
{- i G_{\mu \nu} \over Q^2} \bar u(q_1) ( - i e \gamma_\nu ) v(q_2) \,,
\label{M}
\end{equation}
where $k$, $k^\prime$, $q_1$, $q_2$, and $Q= q_1 + q_2$ are the momenta of the
virtual quark, the final-state quark, the positive lepton, the negative lepton,
and the lepton pair, respectively.  The factor $\Gamma$ is the $4 \times 1$
Dirac spinor associated with the blob in Fig.~\ref{fig:operational}, 
from which the virtual quark
emerges.  If we choose a covariant gauge for QED, the numerator factor in the
photon propagator reduces to $G_{\mu \nu} = g_{\mu \nu}$.  However, if we
square the matrix element (\ref{M}) and integrate over phase space, isolating
those terms with the proper scaling behavior at large $Q_T$, we will not get
the correct fragmentation function.  The reason is that in a covariant gauge,
there are diagrams that do not have the topology of 
Fig.~\ref{fig:operational}, but still
contribute to the scaling part of the cross section.  In order to extract the
complete scaling contribution, it is necessary to use a light-cone gauge in
which the numerator of the photon propagator has the form
\begin{equation}
G_{\mu \nu} \;=\; 
g_{\mu \nu} - {Q_\mu n_\nu + n_\mu Q_\nu \over n \!\cdot\! Q} \,,
\label{prop}
\end{equation}
where $n$ is the light-like vector that defines the longitudinal momentum
fraction of the $\ell^+ \ell^-$ pair:  $z = n \cdot Q / n \cdot k$.  
Note that the $n_\mu Q_\nu$ term in (\ref {prop}) does not contribute 
to the matrix element (\ref{M}), because the leptons are on-shell.  
However, the $Q_\mu n_\nu$ term
does contribute, because the decaying quark is virtual.
The calculations of the fragmentation functions for $c \to \eta_c, J/\psi$ 
in Ref.~\cite{B-C-Y} 
involved diagrams with the topology of Fig.~\ref{fig:operational}, 
except they had a virtual gluon instead of a virtual photon.  
The authors calculated the diagram using the light-cone 
gauge propagator for the gluon.  
In a calculation using the formal definition, 
the contributions corresponding to the additional terms 
in the light-cone gauge propagator come from diagrams 
in which the virtual gluon is emitted by a gluon field from the
eikonal factor (\ref{eikonal-1}).

\begin{figure}
\begin{center}
\begin{tabular}{cc}
\multicolumn{2}{c}{\epsfig{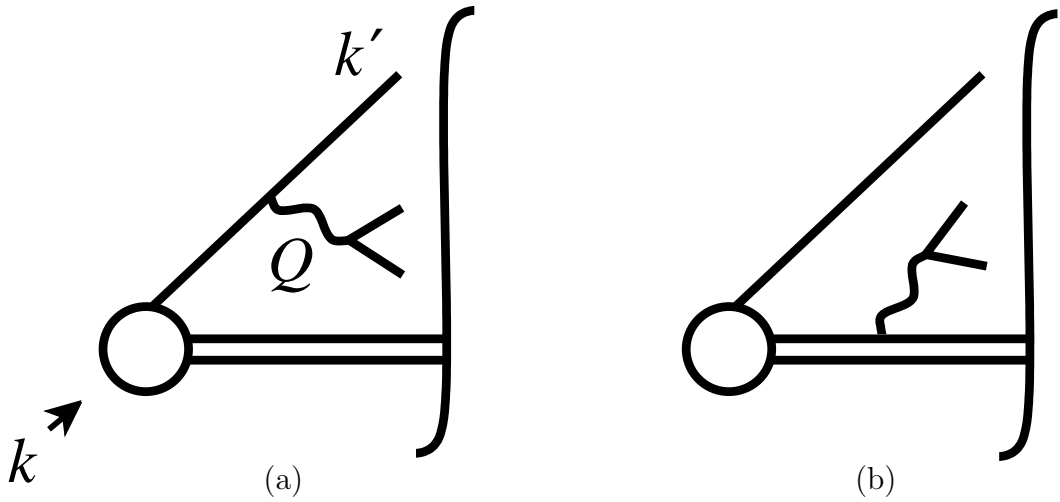}}\\
\\[-5.0ex]
 \mbox{\hspace{4.3cm}}(a)\mbox{\hspace{4cm}}
&\mbox{\hspace{3.2cm}}(b)\mbox{\hspace{3cm}}
\end{tabular}
\end{center}
\caption{
Feynman diagrams for the fragmentation function 
$D_{q \to\ell^+ \ell^-}$
at leading order in $\alpha_s$. 
}
\label{fig:formal}
\end{figure}

If the fragmentation function for a lepton pair is calculated using the
formal definition, it is necessary to generalize the QCD eikonal factor in
(\ref{eikonal-1}) to include a phase from the photon field.  In the
fragmentation function for a quark with electric charge $e_q$, the eikonal
factor becomes
\begin{equation}
{\cal E} = {\cal P} \exp 
\left( i g \int dl \, n^\mu A_\mu^a T^a 
+ i e_q e \int dl \, n^\mu A_\mu \right) \,.
\label{eikonal-2}
\end{equation}
At leading order in $\alpha_s$, the $q \to \ell^+ \ell^-$ 
fragmentation function is given by the square of the sum of the
Feynman diagrams in  Fig.~\ref{fig:formal}. 
The circles represent quark operators, and
the double lines represent the eikonal factor.
A convenient set of Feynman rules for calculating fragmentation 
functions is given in Ref. \cite{Collins-Soper}. 
The contributions that correspond to the additional term 
in the light-cone gauge propagator (\ref{prop}) for the photon 
are provided by the diagram in Fig.~\ref{fig:formal}b in which the virtual
photon is emitted by a photon field from the eikonal factor (\ref{eikonal-2}).

In the calculation of the fragmentation function
for $q \to \ell^+ \ell^-$, the integral over the relative momenta 
of the lepton pair for fixed total momentum $Q$ gives the tensor
$- g^{\mu \nu} + Q^\mu Q^\nu/Q^2$.
In Ref.~\cite{Qiu-Zhang}, Qiu and Zhang decomposed this tensor into
two terms that correspond to the polarization tensors for virtual photons 
that are transversely and longitudinally polarized with respect 
to the direction of the momentum of the lepton pair in a particular frame:
\begin{equation}
- g^{\mu \nu} + {Q^\mu Q^\nu \over Q^2} \;=\;
\left( - g^{\mu \nu} + {n^\mu \bar n^\nu + \bar n^\mu n^\nu \over 2} \right)
+ \left( {Q^\mu Q^\nu \over Q^2} 
	- {n^\mu \bar n^\nu + \bar n^\mu n^\nu \over 2}\right) \,,
\label{tensor}
\end{equation}
where $\bar n$ is a conjugate light-like vector 
satisfying $n \cdot \bar n = 2$.
The frame chosen in Ref.~\cite{Qiu-Zhang}
is the one in which the perpendicular momenta of the  
decaying quark and the lepton pair are $k_\perp' = - Q_\perp/z$ 
and $Q'_\perp =0$. 
In this frame, the second tensor on the right side of (\ref{tensor})
can be expressed as $\epsilon_L^\mu \epsilon_L^\nu$, where 
the polarization vector $\epsilon_L^\mu$ is proportional to
$(n \cdot Q)^2 \bar n^\mu - Q^2 n^\mu$.
This choice for the decomposition into transverse and longitudinal
contributions is rather artificial, but we will adopt it in order 
to compare our results with those of Ref.~\cite{Qiu-Zhang}.

We proceed to calculate the fragmentation functions from the 
square of the sum of the two Feynman diagrams 
in Fig.~\ref{fig:formal}. 
We use dimensional regularization in $4-2 \epsilon$ 
space-time dimensions to regularize ultraviolet divergences.
The fragmentation function can be expressed as an integral
over the invariant mass $Q^2$ of the lepton pair:
\begin{eqnarray}
D^{(0)}_{q \to \ell^+ \ell^-}(z) \;=\;
{C e_q^2 \alpha^2 \over 6 \pi^2}
\int{dQ^2 \over Q^2}
\left({4 \pi \mu^2 \over Q^2}\right)^{2\epsilon}
\left\{ {2(1-z) + (1 - \epsilon) z^2 \over z^2} 
        \left[ z f_1(z) - f_2(z) \right] 
\right.
\nonumber
\\
\left.
+ {2(1-z) \over z^2} f_2(z) \right\} \,,
\label{Dq-0}
\end{eqnarray}
where $C$ is a constant that reduces to 1 as $\epsilon \to 0$:
\begin{eqnarray}
C &=& 
{(1-\epsilon) 2^{2\epsilon} \Gamma({1\over 2}) 
	\over (1-2\epsilon) (1-{2\over 3}\epsilon)
		\Gamma({1\over 2}-\epsilon) } \,.
\label{C}
\end{eqnarray}
In (\ref{Dq-0}), there is an implicit lower limit on $Q^2$ 
that is large enough 
that a perturbative calculation of the QCD corrections 
would be  reliable.
The two terms in the integrand correspond to 
transversely and longitudinally polarized photons, respectively.
The functions $f_n(z)$ can be expressed as integrals over the 
perpendicular components ${\bf Q}_\perp$ of the momentum of the 
lepton pair.  Equivalently, they can be expressed as integrals 
over the invariant mass $s$ of the decaying quark, which is 
related to ${\bf Q}_\perp$ by light-cone energy conservation:
\begin{eqnarray}
s \;=\; {Q^2+Q_\perp^2 \over z} + {Q_\perp^2 \over 1-z} \,.
\label{invmass}
\end{eqnarray}
This implies the lower limit $s > Q^2/z$.
The functions $f_n(z)$ in (\ref{Dq-0}) are given by
\begin{eqnarray}
f_n(z) &=& 
{1 \over \Gamma(1-\epsilon)} [z(1-z)]^{-\epsilon}
\int_{1/z}^\infty dy \, {(y-1/z)^{-\epsilon} \over y^n} \,,
\label{fn-def}
\end{eqnarray}
where $y =s/Q^2$.  They can be evaluated analytically:
\begin{eqnarray}
f_n(z) &=& 
{\Gamma(n-1+\epsilon) \over \Gamma(n)} z^{n-1}(1-z)^{-\epsilon} \,.
\label{fn-analytic}
\end{eqnarray}
The function $f_1(z)$ in (\ref{Dq-0}) has a pole in $\epsilon$.
This ultraviolet divergence is cancelled by the renormalization 
of the composite operator in the definition of the fragmentation 
function.  The renormalized fragmentation function in the 
$\overline{MS}$ scheme is
\begin{eqnarray}
D^{\overline{MS}}_{q \to \ell^+\ell^-}(z,\mu) \;=\;
D^{(0)}_{q \to \ell^+\ell^-}(z)
- {( 4\pi e^{-\gamma} )^\epsilon \over \epsilon}  \frac{e_q^2 \alpha}{2\pi} 
	\int_z^1 {dy \over y} P_{q \to \gamma}(z/y) 
	D_{\gamma \to \ell^+\ell^-}(y) \,,
\label{Dq-ren}
\end{eqnarray}
where $P_{q \to \gamma}$ is the splitting function 
\begin{eqnarray}
P_{q\to \gamma} \;=\; {1+(1-z)^2 \over z} \,,
\end{eqnarray}
and $D_{\gamma \to \ell^+\ell^-}$ is the fragmentation function 
for a photon to decay into a lepton pair:
\begin{eqnarray}
D_{\gamma\to e^+e^-}(z) \;=\;
{C \alpha \over 3 \pi} \int{dQ^2 \over Q^2} 
\left({4 \pi \mu^2 \over Q^2}\right)^\epsilon \, \delta(1-z) \,.
\end{eqnarray}
The coefficient $C$ is given in (\ref{C}).
Our final result for the $\overline{MS}$ fragmentation function is
obtained by taking the limit $\epsilon \to 0$ in (\ref{Dq-ren}): 
\begin{eqnarray}
D^{\overline{MS}}_{q \to \ell^+\ell^-}(z,\mu) \;=\;
{e_q^2 \alpha^2 \over 6 \pi^2} \int{dQ^2 \over Q^2}
\left\{ \left[{1+(1-z)^2 \over z}
        \left(\ln {\mu^2 \over (1-z)Q^2} - 1 \right) -z \right]
+ {2 (1-z) \over z} \right\} \,.
\label{Dq-dr}
\end{eqnarray}
The two terms in the integrand correspond to 
transversely and longitudinally polarized photons, respectively.

In Ref.~\cite{Qiu-Zhang}, Qiu and Zhang defined their
fragmentation functions by imposing the constraint $s < \mu_F^2$
on the invariant mass of the decaying quark, which sets 
the upper limit $y< \mu_F^2/Q^2$ on the integral in 
(\ref{fn-def}).  This eliminates ultraviolet divergences, 
so we can set $\epsilon=0$. 
The results for the integrals are then 
\begin{eqnarray}
f_0(z) &=& 
{1\over z} \left( {z \mu_F^2 \over Q^2} - 1 \right) 
	\theta(\mu_F^2 - Q^2/z) \,,
\\
f_1(z) &=& 
\ln {z \mu_F^2 \over Q^2} \, \theta(\mu_F^2 - Q^2/z) \,,
\\
f_2(z) &=& 
z \left( 1- {Q^2 \over z \mu_F^2} \right)
	\theta(\mu_F^2 - Q^2/z) \,.
\label{f2}
\end{eqnarray}
The resulting expression for the fragmentation function is
\begin{eqnarray}
D_{q \to \ell^+\ell^-}(z,\mu_F) \;=\;
{e_q^2 \alpha^2 \over 6 \pi^2} 
\int{dQ^2 \over Q^2} \, \theta(\mu_F^2 - Q^2/z)
\left\{ {1+(1-z)^2 \over z} \left( \ln {z \mu_F^2 \over Q^2}
                                  - 1 + {Q^2\over z \mu_F^2} \right)
\right.
\nonumber\\
\left.
        + {2(1-z)\over z} \left( 1- {Q^2 \over z \mu_F^2} \right)
        \right\}  \,.
\label{Dq-im}
\end{eqnarray}
This result has been confirmed by Qiu and Zhang \cite{Qiu-Zhang}.
If we take the formal limit $\mu_F^2 \gg Q^2/z$ in (\ref{Dq-im}),
the $\theta$ function becomes 1 and the resulting expression 
differs from (\ref{Dq-dr}) only 
in the argument of the logarithm and in the coefficient of the $z$ term.
The arguments of the logarithm are the same if we make the 
identification $\mu_F^2 = \mu^2/(z(1-z))$.
This can be understood by examining the expression (\ref{invmass})
for the invariant mass of the decaying virtual quark.
It suggests that the scale $\mu$ of dimensional regularization
should be identified not with a cutoff on the invariant mass
of the virtual quark, but with a cutoff on the 
perpendicular momentum $Q_\perp$ of the lepton pair. 
The difference between the coefficients of the $z$ term 
comes from the $\epsilon z f_1(z)$ term in (\ref{Dq-0}), 
which reduces to $z$ in the limit $\epsilon \to 0$.
On the other hand, if we impose an invariant-mass cutoff $\mu_F$
and set $\epsilon=0$, this term vanishes.
Taking the limit $\mu_F \gg Q$ with $z$ fixed in (\ref{Dq-im})
corresponds simply to an alternative ultraviolet cutoff.
The fragmentation function should then differ from (\ref{Dq-dr})
by a finite renormalization of the composite operator 
in the definition of the fragmentation function.
This finite renormalization corresponds to adding the term $-\epsilon z$ 
to the splitting function $P_{q\to \gamma}$ in (\ref{Dq-ren}).

\begin{figure}
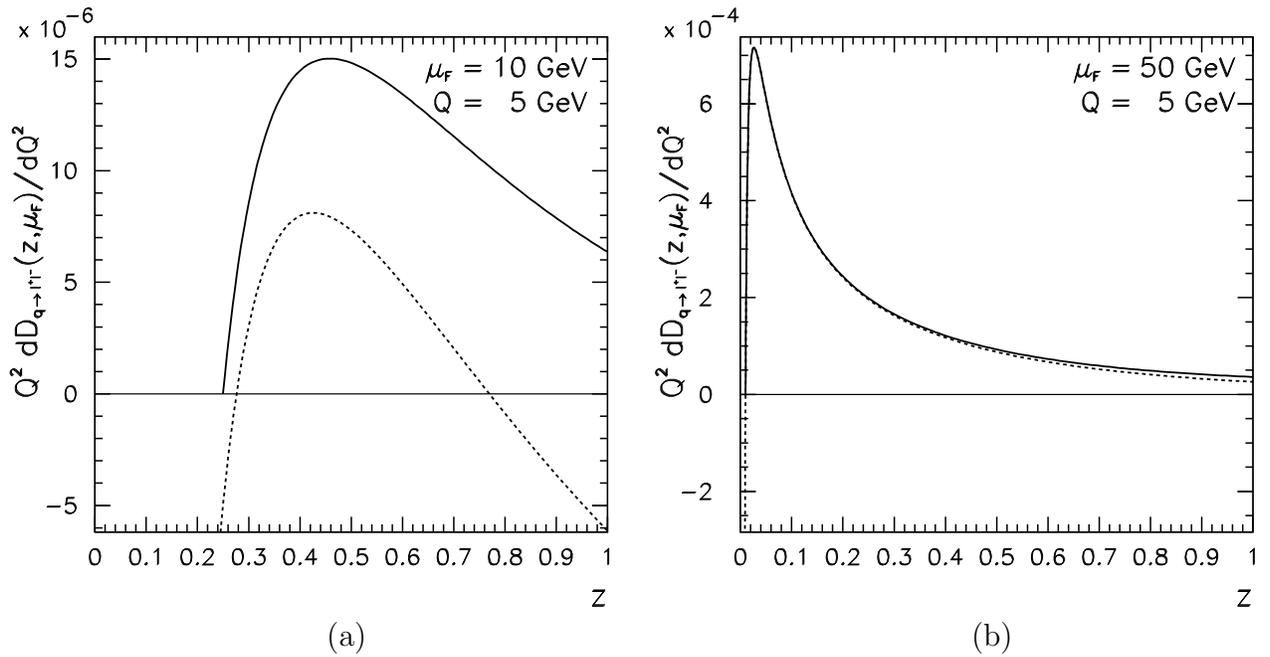

\begin{center}
\begin{tabular}{ccc}
\epsfig{file=lepfrag_fig3a.epsi,width=8cm}
&
&
\epsfig{file=lepfrag_fig3b.epsi,width=8cm}
\\
\mbox{\hspace{1.0cm}}(a)&\mbox{\hspace{0.3cm}}&
\mbox{\hspace{1.0cm}}(b)
\end{tabular}
\end{center}
\caption{
The differential fragmentation function $Q^2 \, dD(z)/dQ^2$
for $q \to\ell^+ \ell^-$ for a lepton pair with invariant mass $Q = 5$ GeV
as a function of $z$ for 
(a)  $\mu_F = 10$ GeV and (b) $\mu_F = 50$ GeV.
The solid curves are the fragmentation functions (\ref{Dq-im})
defined by the invariant-mass cutoff $\mu_F$.
The dashed curves are the fragmentation functions (\ref{Dq-dr})
defined by dimensional regularization with renormalization scale 
$\mu^2 = z(1-z)\mu_F^2$.
}
\label{fig:drim}
\end{figure}

One advantage of the fragmentation function (\ref{Dq-im})
defined by an upper limit on the invariant mass 
is that it builds in threshold effects and the constraints of energy
conservation associated with the decay of a virtual quark with 
invariant mass $\mu_F$.  This might be useful for quantitative 
applications of the fragmentation function. 
In Fig.~\ref{fig:drim}, we compare the differential fragmentation
functions $Q^2 dD(z)/dQ^2$ corresponding to (\ref{Dq-im})
with invariant mass cutoff $\mu_F$ and (\ref{Dq-dr})
with renormalization scale $\mu^2 = z(1-z) \mu_F^2$.
We set $e_q=+{2\over3}$ and $\alpha = {1\over137}$.
We choose $Q=5$ GeV and consider 2 values of $\mu_F$:
$\mu_F = 10$ GeV in Fig.~\ref{fig:drim}a and 
$\mu_F = 50$ GeV in Fig.~\ref{fig:drim}b.
The fragmentation function defined by $s < \mu_F^2$ is 
0 below the threshold at $z=Q^2/\mu_F^2$
and is positive for larger values of $z$. 
The fragmentation function defined by dimensional regularization 
has unphysical negative values over part of its range
as can be seen in Fig.~\ref{fig:drim}.
As $z\to 1$, it remains positive only if $\mu_F > \exp(1) Q$.
As $z \to 0$, it is negative for all $\mu_F$, diverging like 
$\ln(z \mu_F^2/Q^2)/z$.  For large enough values of $\mu_F/Q$ 
as in Fig.~\ref{fig:drim}b, the two fragmentation functions
look similar except that one vanishes for $z<Q^2/\mu_F^2$
and the other becomes negative in that region.
If we keep $\mu$ fixed in (\ref{Dq-dr}) instead of $\mu_F$,
the fragmentation function still exhibits unphysical behavior.
It is positive-definite if $\mu>Q$,
but it diverges like $\ln(\mu^2/Q^2)/z$ as $z\to0$  
and like $\ln(\mu^2/((1-z)Q^2))$ as $z\to 1$.
We conclude that the fragmentation function (\ref{Dq-dr})
defined by dimensional regularization is of little practical use.
It is essential to take into account threshold effects 
in some way, such as by imposing an 
upper limit on the invariant mass as in (\ref{Dq-im}).

If the QED phase in the eikonal factor (\ref{eikonal-2}) 
is omitted in the formal definition of the
fragmentation function, 
$D^{(0)}_{q \to \ell^+ \ell^-}$ is given by the square 
of the Feynman diagram in Fig.~\ref{fig:formal}a.
The resulting expression is not gauge invariant,
but it is independent of the gauge parameter for covariant gauges.
Using dimensional regularization, we obtain
\begin{eqnarray}
D^{(0)}_{q \to \ell^+ \ell^-}(z) \;=\;
{C e_q^2 \alpha^2 \over 6 \pi^2}
\int{dQ^2 \over Q^2}
\left({4 \pi \mu^2 \over Q^2}\right)^{2\epsilon}
\left\{ {2(1-z) + (1 - \epsilon) z^2 \over z^2} 
		\left[ z f_1(z) - f_2(z) \right] 
\right.
\nonumber
\\
\left. + {1-z \over 2z^2} \left[ z^2 f_0(z) - 4 z f_1(z) + 4 f_2(z) \right] 
	\right\} \,.
\label{Dq-dr:QZ}
\end{eqnarray}
The two terms in the integrand correspond to 
transversely and longitudinally polarized photons, respectively.
The transverse term is identical to that in (\ref{Dq-dr}).
This follows from the fact that the Feynman rule for the 
emission of a virtual photon from the eikonal line 
in Fig.~\ref{fig:formal}b is proportional to $n^\mu$, 
which is orthogonal to the transverse tensor in (\ref{tensor}).
Thus the omission of the diagram in Fig.~\ref{fig:formal}b 
can only affect the longitudinal term.
With dimensional regularization, the function $f_0(z)$ vanishes, 
as is evident from (\ref{fn-analytic}).
The function $f_1(z)$ has a pole in $\epsilon$.
In the transverse term, the pole can be removed by the 
renormalization (\ref{Dq-ren}).  However, there is also a pole in the
longitudinal term that is not removed by renormalization of the 
composite operator.  This failure of renormalization is the signal
that the definition of the fragmentation function that omits 
the QED phase in the eikonal factor is inconsistent.

\begin{figure}
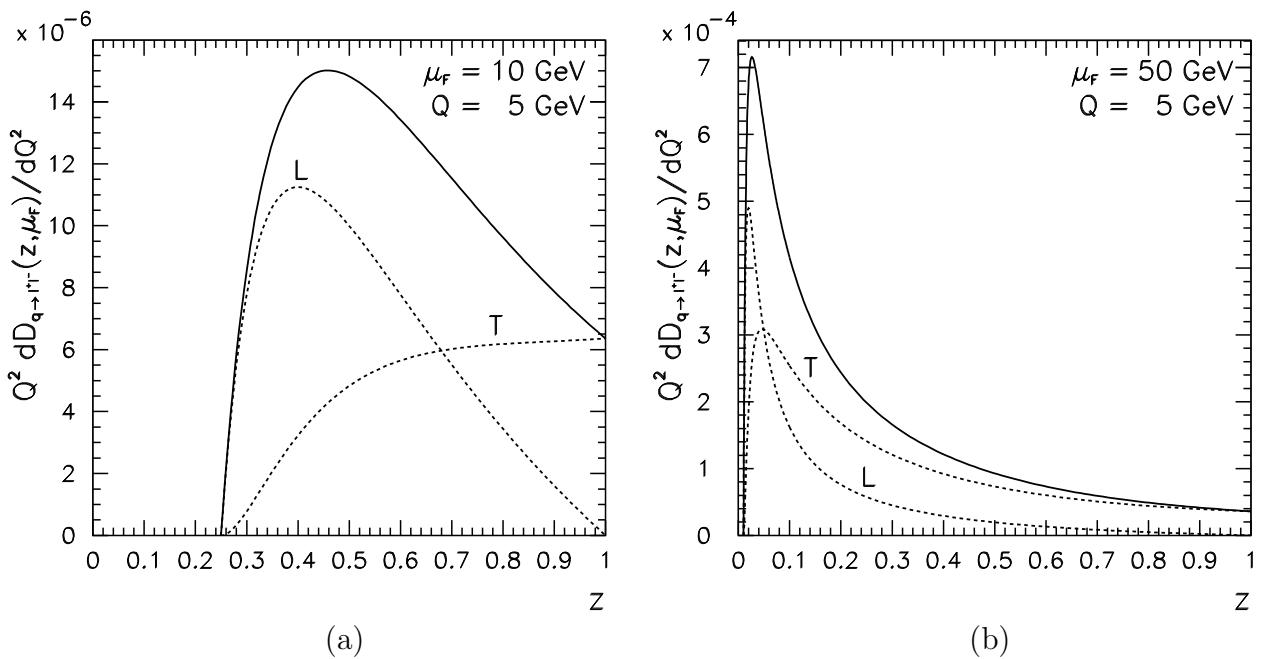

\begin{center}
\begin{tabular}{ccc}
\epsfig{file=lepfrag_fig4a.epsi,width=8cm}
&
&
\epsfig{file=lepfrag_fig4b.epsi,width=8cm}
\\
\mbox{\hspace{1.0cm}}(a)&\mbox{\hspace{0.3cm}}&
\mbox{\hspace{1.0cm}}(b)
\end{tabular}
\end{center}
\caption{
The differential fragmentation function $Q^2 \, dD(z)/dQ^2$
for $q \to\ell^+ \ell^-$ for a lepton pair with invariant mass $Q = 5$ GeV
as a function of $z$ for the invariant-mass cutoffs
(a) $\mu_F = 10$ GeV and (b) $\mu_F = 50$ GeV.
The solid curve is the total fragmentation function.
The dashed curves labelled $T$ and $L$ are the contributions from
transverse and longitudinal virtual photons.
}
\label{fig:fragfun}
\end{figure}

In Fig.~\ref{fig:fragfun}, we compare the transverse and longitudinal
contributions to the fragmentation function 
(\ref{Dq-im}) calculated using an invariant-mass cutoff.
We choose $Q=5$ GeV and consider 2 values of $\mu_F$:
$\mu_F = 10$ GeV in Fig.~\ref{fig:fragfun}a and $\mu_F = 50$ GeV 
in Fig.~\ref{fig:fragfun}b.
The dashed curves labelled $T$ and $L$ are the transverse and longitudinal
contributions given by the two terms in (\ref{Dq-im}).
Their sum is the solid curve.
The longitudinal polarization dominates just above the threshold
at $z = Q^2/\mu_F^2$, 
because the longitudinal term increases linearly in $z - Q^2/\mu_F^2$
while the transverse term increases quadratically.
The transverse polarization dominates at large $z$,
because the longitudinal term vanishes at $z=1$.

We can define a longitudinal polarization fraction $\xi$
by dividing the longitudinal fragmentation probability 
by the total fragmentation probability for lepton pairs 
of invariant mass $Q^2$.  The fragmentation probability 
is proportional to the integral over $z$ of $dD(z)/dQ^2$.
The longitudinal polarization fraction for $Q = 5$ GeV
decreases from $\xi = 0.56$ at $\mu_F = 10$ GeV to 0.34
at $\mu_F = 50$ GeV and to 0.23 at $\mu_F = 250$ GeV.
This is in accord with the intuition that the virtual photon should 
behave more and more like a real photon as $\mu_F$ increases.

The longitudinal polarization fraction 
defined by the ratio of fragmentation probabilities 
decreases rather slowly with $\mu_F$.
However a more relevant measure of the polarization
is the ratio of the second moments of the fragmentation functions.
The reason is that the hard-scattering cross section $d \hat \sigma$
scales like $1/(Q_T/z)^2$, and this weights the fragmentation 
function by $z^2$.  
The longitudinal polarization fraction for $Q = 5$ GeV
defined  by the ratio of the second moments
decreases from $\xi = 0.43$ at $\mu_F = 10$ GeV to 0.15
at $\mu_F = 50$ GeV and to 0.08 at $\mu_F = 250$ GeV.

An important observable in lepton pair production is the 
angular distribution of the momentum of one of the leptons
in the rest frame of the lepton pair. 
The angular distribution is proportional to $1 + \alpha \cos^2 \theta$, 
where $\theta$ is the angle between the momentum ${\bf q}_1$ 
of the negative lepton and some quantization axis.
The polarization variable $\alpha$ is related to the fraction $\xi$ 
of lepton pairs that come from longitudinally polarized virtual photons
by $\alpha= (1-3\xi)/(1+\xi)$.
The fraction $\xi$ depends on the choice of quantization axis.
The choice adopted in Ref.~\cite{Qiu-Zhang} is rather artificial,
because it requires specifying the transverse momentum 
of the fragmenting quark which is not easily observed.
In the case of a hadron collider, a more physical choice 
for the quantization axis is the direction of the momentum ${\bf Q}$ 
of the lepton pair in the rest frame of the colliding hadrons.
In this case, the longitudinal polarization vector $\epsilon_L^\mu$ 
is a linear combination of $Q$ 
and the total momentum $K$ of the colliding hadrons:
\begin{equation}
\epsilon_L^\mu \;=\;
{ Q^2 K^\mu - (K \!\cdot\! Q) Q^\mu
\over [(K \!\cdot\! Q)^2- K^2 Q^2]^{1/2} [Q^2]^{1/2} } \,.
\label{eps-L}
\end{equation}
The contribution to the fragmentation function from 
longitudinally polarized virtual photons can be obtained by replacing
the lepton tensor $- g^{\mu \nu} + Q^\mu Q^\nu/Q^2$
by $\epsilon_L^\mu \epsilon_L^\nu$.
The resulting expression for the fragmentation function 
depends explicitly on $k_+ = k \cdot n$:
\begin{eqnarray}
D_{q \to (\ell^+ \ell^-)_L}(z,\mu_F) \;=\;
{e_q^2 \alpha^2 \over 6 \pi^2}
\int{dQ^2 \over Q^2}
{2(1-z) \over z^2} f_2(z,\mu_F/Q,Q/k_+) \,,
\end{eqnarray}
where $f_2$ is a function of $z$, $\mu_F/Q$, and $Q/k_+$:
\begin{eqnarray}
f_2(z,\mu_F/Q,Q/k_+) &=& 
\int_{1/z}^{\mu_F^2/Q^2} {dy  \over y^2} 
{ [z-y(Q/k_+)^2]^2 \over [z +(1+y-yz)(Q/k_+)^2]^2 - 4(Q/k_+)^2}\,.
\end{eqnarray}
In the limit $k_+ \gg Q$, this reduces to (\ref{f2}).
The complete fragmentation function summed over polarizations 
is independent of $Q/k_+$ and is given by (\ref{Dq-im}):
\begin{eqnarray}
D_{q \to \ell^+\ell^-}(z,\mu_F) \;=\;
{e_q^2 \alpha^2 \over 6 \pi^2} 
\int{dQ^2 \over Q^2} \, \theta(\mu_F^2 - Q^2/z)
\left\{ {1+(1-z)^2 \over z} \ln {z \mu_F^2 \over Q^2}
        - z \left( 1 - {Q^2\over z \mu_F^2} \right)
        \right\}  \,.
\end{eqnarray}
In the frame defined by ${\bf K} = 0$, 
$k_+$ is the sum of the energy and momentum of an on-shell quark
produced by some hard scattering.
A reasonable choice for $\mu_F$ is the transverse momentum
of that quark, give or take a factor of 2.

The NRQCD factorization approach predicts that the $1^{--}$ quarkonium 
states should become increasingly transversely polarized
as their transverse momentum $P_T$ increases 
\cite{Cho-Wise,Beneke-Rothstein}.
Quantitative predictions of the polarization of the $\psi(2S)$ 
\cite{Beneke-Kramer,Leibovich}, $J/\psi$ \cite{B-K-L},
and $\Upsilon(2S)$ \cite{Braaten-Lee:Ups}
indicate that the increase in the polarization 
should set in at values of $P_T$ that are accessible at the Tevatron.
The present data on the polarization of $J/\psi$ and $\psi(2S)$ 
from the CDF collaboration \cite{CDF} seem to indicate a decrease
in the transverse polarization at large $P_T$, although in both cases the
discrepancy with the prediction is significant only in the largest $P_T$ bin.
The argument that the tranverse polarization of $J/\psi$ or $\psi(2S)$ 
should increase with $P_T$ is completely analogous to the corresponding
argument for lepton pairs, except that it involves a virtual gluon 
instead of a virtual photon.  There are many effects that could dilute the
transverse polarization or delay the onset of the predicted increase.
However no plausible mechanisms have been identified
that could make it decrease with $P_T$.  
We expect that more accurate measurements from Run II of the 
Tevatron will reveal the increase in transverse polarization 
predicted by NRQCD.

In conclusion, we have calculated the fragmentation function 
for a light quark to decay into a lepton pair 
to leading order in $\alpha_s$.
For renormalizability and for QED gauge invariance,
it is essential to include a QED phase in the eikonal factor 
in the formal definition of the fragmentation function.
Berger, Gordon, and Klasen \cite{B-G-K} have shown that the 
distribution of the transverse momentum $Q_T$ of lepton pairs 
in hadron collisions is dominated by parton processes 
initiated by gluons if $Q_T > Q/2$.
The $Q_T$ distribution can therefore provide useful constraints 
on the parton distribution for gluons.
Our fragmentation function for $q \to \ell^+ \ell^-$ may be useful for
calculating the $Q_T$ distribution in the limit $Q_T \gg Q$.

J.L. thanks the IPPP at University of Durham
and the High Energy Theory Group at Ohio State University
for their hospitality while this work was being carried out.
This work was supported in part by the U.S. Department of Energy 
under grant DE-FG02-91-ER40690
and by the KOSEF and the DFG through
the German-Korean scientific exchange program DFG-446-KOR-113/137/0-1.
We thank J.-W. Qiu for pointing out an error in the normalization 
of our fragmentation functions in a previous version of this paper.


\newpage



\begin{references}
\bibitem{theorems}
J. C. Collins and D. E. Soper, 
	Ann. Rev. Nucl. Part. Sci. {\bf 37}, 383 (1987).
\bibitem{Curci-Furmanski-Petronzio}
G. Curci, W. Furmanski, and R. Petronzio,
Nucl. Phys. B {\bf 175}, 27 (1980). 
\bibitem{Collins-Soper}
J. C. Collins and D. E. Soper, Nucl. Phys. B {\bf 194}, 445 (1982). 
\bibitem{Kniehl-Kramer-Poetter}
B. A. Kniehl, G. Kramer, and B. P\"{o}tter,
Phys. Rev. Lett. {\bf 85}, 5288 (2000).
\bibitem{B-B-L}
G. T. Bodwin, E. Braaten, and G. P. Lepage,
Phys.\ Rev.\ D {\bf 51}, 1125 (1995); {\bf 55}, 5855(E) (1997).
\bibitem{Braaten-Yuan}
E. Braaten and T. C. Yuan, Phys. Rev. Lett. {\bf 71}, 1673 (1993);
Phys. Rev. D {\bf 52}, 6627 (1995).
\bibitem{B-C-Y}
E. Braaten, K. Cheung, and T. C. Yuan,
Phys. Rev. D {\bf 48}, 4230 (1993).
\bibitem{Ma}
J. P. Ma,
Phys. Lett. B {\bf 332}, 398 (1994). 
\bibitem{Braaten-Lee:NLO}
E. Braaten and J. Lee, Nucl. Phys. B {\bf 586}, 427 (2000);   
J. Lee, hep-ph/0009111.
\bibitem{Qiu-Zhang}
J.-W. Qiu and X.-F. Zhang, hep-ph/0101004. 
\bibitem{Cho-Wise}
P. Cho and M. B. Wise, Phys. Lett. B {\bf 346}, 129 (1995). 
\bibitem{Beneke-Rothstein}
M. Beneke and I.Z. Rothstein, Phys. Lett. B {\bf 372}, 157 (1996); 
	{\bf 389}, 769(E) (1996). 
\bibitem{Beneke-Kramer}
M. Beneke and M. Kr\"amer, Phys. Rev. D {\bf 55}, 5269 (1997). 
\bibitem{Leibovich}
A. K. Leibovich, Phys. Rev. D {\bf 56}, 4412 (1997). 
\bibitem{B-K-L}
E. Braaten, B. A. Kniehl, and J. Lee, Phys. Rev. D {\bf 62}, 094005 (2000). 
\bibitem{Braaten-Lee:Ups}
E. Braaten and J. Lee, hep-ph/0012244. 
\bibitem{CDF}
The CDF Collaboration, 
T. Affolder {\it et al.}, 
	Phys. Rev. Lett. {\bf 85}, 2886 (2000). 
\bibitem{B-G-K}
E. L. Berger, L. E. Gordon, and M. Klasen,
Phys. Rev. D {\bf 58}, 074012 (1998).
\end{references}
\end{document}